\documentclass[conference,10pt,a4paper]{IEEEtran}
\usepackage{amsmath,amssymb,amsfonts}
\usepackage{textcomp}
\def\BibTeX{{\rm B\kern-.05em{\sc i\kern-.025em b}\kern-.08em
T\kern-.1667em\lower.7ex\hbox{E}\kern-.125emX}}

\usepackage{subfigure}
\usepackage[pdftex]{graphicx}
\usepackage{caption}  
\usepackage{subcaption}
\graphicspath{ {./img/} }

\usepackage{cite}
\usepackage{makeidx}

\usepackage{algorithm}
\usepackage{algorithmic}

\usepackage{multirow}
\usepackage{hyperref} 
\usepackage{cleveref} 

\hypersetup{colorlinks = true, 
	    linkcolor = black, 
	    urlcolor = black,
        citecolor = black} 

\begin{document}

\title{
Computational Imaging-Based ISAC Method with Large Pixel Division
}
\author{Xin~Tong\textsuperscript{$\dagger$}\textsuperscript{$\ddagger$},
        Zhaoyang~Zhang\textsuperscript{$\dagger$},
        Zhaohui~Yang\textsuperscript{$\dagger$},
        Yu~Ge\textsuperscript{$\ddagger$},
        Henk~Wymeersch\textsuperscript{$\ddagger$}\\
        \textsuperscript{$\dagger$}College of Information Science and Electronic Engineering, Zhejiang University, Hangzhou, China\\
        \textsuperscript{$\dagger$}Zhejiang Provincial Key Laboratory of Info. Proc., Commun. \& Netw. (IPCAN), Hangzhou, China\\
        \textsuperscript{$\ddagger$}Department of Electrical Engineering, Chalmers University of Technology, Gothenburg, Sweden\\
        E-mails: \{tongx, ning\_ming, yang\_zhaohui\}@zju.edu.cn, \{yuge, henkw\}@chalmers.se
        
        \thanks{This work was supported in part by National Natural Science Foundation of China under Grants 62394292 and U20A20158, Zhejiang Provincial Key R\&D Program under Grant 2023C01021, Ministry of Industry and Information Technology under Grant TC220H07E, the Fundamental Research Funds for the Central Universities, and the SNS JU project 6G-DISAC under the EU's Horizon Europe research and innovation Program under Grant Agreement No 101139130.}
}

\maketitle

\begin{abstract}
  One of the key points in designing an integrated sensing and communication (ISAC) system using computational imaging is the division size of imaging pixels. If the size is too small, it leads to a high number of pixels that need processing. On the contrary, it usually causes large processing errors since each pixel is no longer uniformly coherent. In this paper, a novel method is proposed to address such a problem in environment sensing in millimeter-wave wireless cellular networks, which effectively cancels the severe errors caused by large pixel division as in conventional computational imaging algorithms. To this end, a novel computational imaging model in an integral form is introduced, which leverages the continuous characteristics of object surfaces in the environment and takes into account the different phases associated with the different parts of the pixel. The proposed algorithm extends computational imaging to large wireless communication scenarios for the first time. The performance of the proposed method is then analyzed, and extensive numerical results verify its effectiveness. 
 \end{abstract}


\IEEEpeerreviewmaketitle

\section{Introduction}
In future, wireless communication technology faces larger-scale environments and more complex applications \cite{Saad, ge2, Wei}. With emerging technologies, such as autonomous driving, unmanned aerial vehicles, intelligent robots, etc., more accurate environmental information is required, including location, shape, and electromagnetic (EM) characteristics of objects in the surrounding environment. Currently, integrated sensing and communication (ISAC) technology \cite{Liu, ge, tongx3}, which aims to utilize ubiquitous wireless signals to achieve environment sensing based on the wireless communication framework, has become a hot research topic.

In wireless communication scenarios, the complex environment will affect wireless channels. One of the significant challenges in ISAC system design is addressing a huge number of unknown environmental factors, including the number, location, and scattering characteristics of objects present in the environment. 
To address this issue, EM imaging technology is introduced into ISAC system design. Recently, there have been many imaging technologies that have the potential to achieve high-precision imaging in wireless communication scenarios, including computational imaging \cite{zhang}, holographic imaging \cite{Torcolacci}, radar coincidence imaging \cite{Zhu}, etc.

The computational imaging method can obtain environmental information through pixel division, that is, dividing the unknown environment into discrete pixels as the smallest imaging unit and imaging by calculating the value of each pixel. Computational imaging has the advantage of enabling imaging with high resolution, especially for tiny targets whose size is close to the wavelength. Consequently, the computational imaging method has many applications for confined imaging regions, including medical examinations and security checks \cite{Sun}. 
Computational imaging technology has recently been applied to wireless communication systems. Based on computational imaging and multiple access technology, a millimeter-wave ISAC system exploiting the sparsity of the environment is proposed in \cite{tongx}. The complex propagation characteristics of wireless signals, occlusion, and diffraction characteristics are jointly considered in computational imaging technology to achieve imaging in wireless communication systems \cite{tongx2}.

However, conventional computational imaging is typically limited to confined imaging regions. There are still many practical issues need to be considered when performing computational imaging in large-scale wireless communication scenarios, especially errors caused by pixel division. To be exact, the pixel cannot represent the EM characteristics of the area where the pixel is located \cite{tongx}. Therefore, how to set the size of pixels to achieve a trade-off between computational efficiency and sensing accuracy has become an unprecedented problem for computational imaging in wireless communication scenarios.

In this paper, we explore the continuous characteristics of object surfaces in the environment and propose a computational imaging model to cancel the phase errors caused by pixel division that significantly contribute to large pixel division errors. In the proposed model, we jointly process the signals of multiple transceivers and carriers and convert the environment sensing problem into a compressed sensing (CS) reconstruction problem \cite{Donoho}, leveraging the sparse characteristics of the environment. By solving the CS problem with a sparse vector reconstruction algorithm, we can obtain the image of the target objects. We also compare and analyze the errors associated with the pixel division method in conventional computational imaging versus those in the proposed method.
The main contributions of this paper are summarized as follows: \textit{(i)}
 We propose an ISAC scheme that leverages the continuous characteristics of
  object surfaces within the environment to enable large-scale environment sensing;
  %
  \textit{(ii)}
  We propose a novel computational imaging model based on the integral form to achieve the cancellation of the phase error, which extends computational imaging to a large wireless communication scenario for the first time;
  %
  \textit{(iii)}
  We analyze the performance of the proposed computational imaging model. Extensive numerical results verify the effectiveness of the proposed method. 

\begin{figure}[t]
  \centering
  \includegraphics[width=0.99\linewidth]{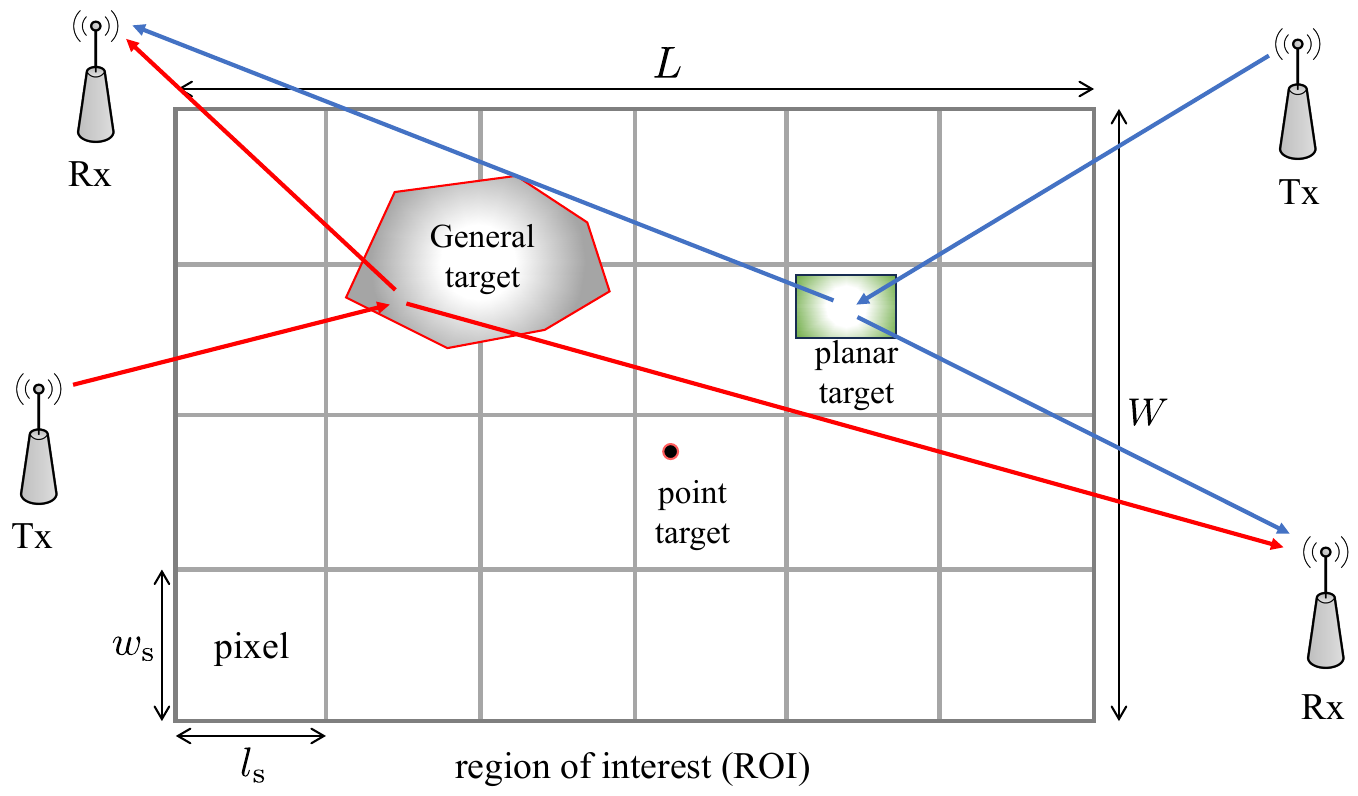}
  \caption{The considered 2D ISAC scenario and pixel division method with multiple Txs and Rxs.
  }
  \label{fig1}
  \end{figure}

\section{Computational Imaging-Based ISAC Setting}
\subsection{Environment Setting}
We consider a simplified 2-dimensional (2D)\footnote{It is worthy noted that the analysis method for 3-dimensional (3D) scenarios can be derived using a similar approach.} wireless communication scenario, as depicted in Fig.~\ref{fig1},  where multiple transmission antennas (Txs) send communication signals to multiple receive antennas (Rxs).  
We consider the synchronization between the transceivers at the symbol level and adopt a specific communication modulation method such as orthogonal frequency division multiplexing (OFDM). We use the channel state information (CSI) to sense the environment.
We assume that the environment sensing between a Tx and an Rx remains stationary within the coherence time. 
Our purpose is to process and detect the environmental information from the communication signals, enabling effective environment sensing within this wireless communication framework.

As shown in Fig.~\ref{fig1}, we define the region of interest (ROI) as the environment to be sensed. Based on the computational imaging method, the 2D environment is uniformly divided into multiple pixels. The length and width of each pixel are $l_{\rm{s}}$ and $w_{\rm{s}}$ respectively, and each pixel is considered as the smallest unit for environment sensing.
Let $L$ and $W$ denote the length and width of the ROI, respectively, thus, the total number of pixels in the ROI is given by $N_{\rm{s}} =\frac{ L}{l_{\rm{s}} }\cdot \frac{W} {w_{ \rm{s}}}$. 
We represent the environmental information in the $n_{\rm{s}}$-th pixel using the scattering coefficient $x_{n_{\rm{s}}}$.
If no target object is present in the $n_{\rm{s}}$-th pixel, then $x_{n_{\rm{s}}} = 0$; otherwise, $x_{n_{\rm{s}}} \in (0,1]$.
Therefore, the environmental information for the ROI can be characterized by the scattering coefficient vector, denoted as ${{\bf x}} = [x_1, x_2, \ldots, x_{N_{\rm{s}}}]^{\rm T} \in \mathbb{R}^{{N_{\rm{s}}} \times 1}$.

\subsection{Traditional Propagation Model}
In existing works \cite{tongx, tongx2}, the environment sensing method based on computational imaging divides the environment into several pixels, obtaining the sensing result by calculating the scatter coefficients for each pixel. To ensure high sensing accuracy, we consider multiple subcarriers. For the $k$-th subcarrier, the communication signals between a pair of transceivers propagate through multipath channels. The multipath propagation gain ${\bf H}_{k}^{\rm NLOS} \in \mathbb{C}^{N_{\rm{R}} \times N_{\rm{T}}}$ consists of two main parts: the free space propagation gain ${\bf{H}}_{k}^{\rm{Tx}}\in \mathbb{C}^{N_{\rm{s}} \times N_{\rm{T}}}$ from the Tx to the environment scatterer, and the free space propagation gain ${\bf{H}}_{k}^{\rm{Rx}}\in \mathbb{C}^{N_{\rm{R}} \times N_{\rm{s}}}$ from the environment scatterer to the Rx, where ${N_{\rm{T}}}$ and ${N_{\rm{R}}}$ are the number of Txs and Rxs, respectively.

In conventional computational imaging methods, the aforementioned free-space propagation gain is calculated through a statistical channel model, based on the antenna and pixel positions. Let the position of ${n_{\rm{T}}}$-th Tx be $(x_{n_{\rm{T}}},y_{n_{\rm{T}}})$ and the position of ${n_{\rm{s}}}$-th pixel center point be $(x_{n_{\rm{s}}},y_{n_{\rm{s}}})$, the free-space propagation gain from the ${n_{\rm{T}}}$-th Tx to the ${n_{\rm{s}}}$-th pixel can be calculated as
\begin{align}
    {\bf{H}}_{k}^{\rm{Tx}}({n_{\rm{s}}},{n_{\rm{T}}})& = \frac{\lambda_{k}}{4\pi d} e^{{-2j\pi  d}/{\lambda_{k}}},
    \label{eq1}
\end{align}
where $d = \sqrt{(x_{n_{\rm{T}}} - x_{n_{\rm{s}}})^2 + (y_{n_{\rm{T}}} - y_{n_{\rm{s}}})^2}$,   $\lambda_{k}$ is the wavelength of the $k$-th subcarrier, and $d$ is the distance between the antenna and the pixel. In addition, ${\bf{H}}_{k}^{\rm{Rx}}$ can be computed in a similar way as  ${\bf{H}}_{k}^{\rm{Tx}}$ in \eqref{eq1}.

It is impractical to directly apply \eqref{eq1} to environment sensing. The primary reason is that the phase errors resulting from pixel division significantly impact the performance of computational imaging. In \eqref{eq1}, a phase error occurs when the target in the environment deviates from the center of the pixel. There are wave path differences between the path from the antenna to the pixel and the path from the antenna to the point target, which introduces the phase error in the propagation gain of the two paths. Therefore, when the pixel size exceeds one wavelength, the phase error becomes almost random. The propagation phase characteristic from the center point of the pixel to the antenna cannot represent the propagation phase characteristics of all locations within this pixel range. A detailed theoretical analysis of these errors is provided in Section \ref{sec4}.

\section{Proposed Model and Algorithm Design}
\subsection{Proposed Model}
The conventional computational imaging method can only support very small pixels, resulting in significant overhead in sensing resources, including the number of antennas, carriers, storage space, computational complexity, etc. To achieve the sensing of the large-scale environment, we propose a method of using large pixels for computational imaging. By considering the continuous characteristics of the object surface, we divide the scenario into pixels while employing an integral method to precisely calculate the propagation gain from the antenna to each pixel position. The impact of the entire pixel area on the propagation gain is considered, and the propagation gain in \eqref{eq1} can be reformulated as\footnote{We  focus on the problem of pixel division error, and other physical characteristics of the scattering coefficient (e.g., variation within a pixel and frequency dependence) will be left for future work.}
\begin{align}
  {\bf{\tilde H}}_{k}^{\rm{Tx}}({n_{\rm{s}}},{n_{\rm{T}}}) &= \int_{x_{n_{\rm{s}}} - l_{\rm s}/2}^{x_{n_{\rm{s}}} + l_{\rm s}/2}\int_{y_{n_{\rm{s}}} - w_{\rm s}/2}^{y_{n_{\rm{s}}} + w_{\rm s}/2} \frac{\lambda_{k}}{4\pi dl_{\rm s}w_{\rm s}} e^{\frac{-2j\pi  d}{\lambda_{k}}} {\rm d}x{\rm d}y,
  \label{eqtx}
\end{align}
where $d = \sqrt{(x_{n_{\rm{T}}} - x)^2 + (y_{n_{\rm{T}}} - y)^2}$. 
In addition, the calculation of ${\bf{\tilde H}}_{k}^{\rm{Rx}}$ can be obtained using a similar method as in \eqref{eqtx}.
Therefore, in the $k$-th subcarrier, the received signal ${\bf Y}_{k} \in \mathbb{C}^{N_{\rm{R}} \times T}$ of $N_{\rm{R}}$ Rxs can be expressed as
\begin{align}\label{eq3}
  {\bf Y}_{k}  &= \big({\bf{\tilde H}}_{k}^{\rm Rx} {\rm diag}({\bf x}) {\bf{\tilde H}}_{k}^{\rm Tx} + {\bf H}_{k}^{\rm LOS} \big){\bf S}_{k} + {\bf N} \nonumber \\
  &= \big({\bf H}_{k}^{\rm NLOS} + {\bf H}_{k}^{\rm LOS} \big){\bf S}_{k} + {\bf N},
\end{align}
where ${\bf H}_{k}^{\rm LOS} \in \mathbb{C}^{N_{\rm{R}} \times N_{\rm{T}}}$ is the line-of-sight (LOS) propagation gain between ${N_{\rm T}}$ Txs and ${N_{\rm R}}$ Rxs, ${\bf S}_{k} \in \mathbb{C}^{N_{\rm{T}} \times T}$ denotes the transmission sequence of ${N_{\rm T}}$ Txs in the $k$-th subcarrier{\footnote{In practical systems, orthogonal pilots can be used as transmission signals for channel estimation like in the following section.}}, $T$ is the sequence length, and ${\bf N} \in \mathbb{C}^{N_{\rm{R}} \times T}$ denotes the noise.

Compared with traditional models, the proposed computational imaging model can mitigate the impact of phase errors. If the pixels are divided on a uniform plane, the error of pixel division can be entirely eliminated. Errors arise only when the pixel area exceeds the target object's area, or when the target object's scattering coefficient is uneven. However, these errors primarily affect amplitude and will be left to future work. The specific theoretical analysis of errors is given in Section IV.

\subsection{Compressed Sensing Reconstruction Formulation}
Environment sensing is achieved by jointly processing the received signals of Rxs. Upon receiving ${\bf Y}_{k}$ by all $N_{\rm{R}}$ Rxs in the $k$-th subcarrier, 
the unknown environmental information $\bf x$ can be estimated through exploiting the known sequences (pilots) ${\bf S}_{k}$ sent by Txs during the communication procedure. 
Since the LOS propagation gain ${\bf H}_{k}^{\rm LOS}$ does not contain any multipath components, it can be accurately estimated with the point-to-point mathematical model as described in \eqref{eq1}. Therefore, the LOS propagation gain ${\bf H}_{k}^{\rm LOS}$ can be effectively cancelled.
Using the transmitted sequence ${\bf S}_{k}$ and the received sequence ${\bf Y}_{k}$, based on the conventional channel estimation method \cite{Beek}, we obtain the estimated propagation gain ${\bf \hat H}_{k}$, which can be expressed as 
\begin{equation}\label{eq4}
  {\bf \hat H}_{k} = {\bf{\tilde H}}_{k}^{\rm Rx} {\rm diag}({\bf x}) {\bf{\tilde H}}_{k}^{\rm Tx} + {\bf \hat N},
\end{equation}
according to \eqref{eq3}, where $\bf \hat N$ is the channel estimation error.

Based on \eqref{eq4}, we jointly process the multipath propagation gain of $K$ subcarriers, which results in
\begin{equation}
    {\bf \hat H} = {\bf A}{\bf x} + \bf{n},  \label{yax}
\end{equation}
where
\begin{equation}
    {\bf \hat H} =\left[
    \begin{array}{c}
        {\bf \hat H}_{1} \\
        \vdots \\
        {\bf \hat H}_{k}\\
        \vdots \\
        {\bf \hat H}_{K}\\
    \end{array}
    \right]_{M \times 1,}  \bf{A}= 
    \left[
    \begin{array}{c}
      {\bf{\tilde H}}_{1}^{\rm Rx} {\rm diag}([{\bf{\tilde H}}_{1}^{\rm Tx}]_{:,1})\\
        \vdots \\
        {\bf{\tilde H}}_{k}^{\rm Rx} {\rm diag}([{\bf{\tilde H}}_{k}^{\rm Tx}]_{:,n_{\rm T}})\\
        \vdots \\
        {\bf{\tilde H}}_{K}^{\rm Rx} {\rm diag}([{\bf{\tilde H}}_{K}^{\rm Tx}]_{:,N_{\rm T}})\\
    \end{array}
    \right]_{{\it M} \times {\it N}_{\rm s},}\nonumber
\end{equation}
with $M=KN_{\rm T}N_{\rm R}$.
Due to the sparsity of environmental scatterers, the environment sensing problem is transformed into a CS reconstruction problem as
\begin{equation}\label{eq6}
  {\bf \hat x} = \arg \min_{\bf x} \|{\bf x}\|_0 \;\; {\rm s.t.}\;\; \|{\bf \hat H}_{k} - {\bf{\tilde H}}_{k}^{\rm Rx} {\rm diag}({\bf x}) {\bf{\tilde H}}_{k}^{\rm Tx}\|_2 \leq \varepsilon,
\end{equation}
where $\varepsilon$ is a slack variable.

\subsection{Algorithm Design}
The CS reconstruction problem in \eqref{eq6} can be solved by 
the generalized approximate message passing (GAMP) algorithm \cite{Rangan}, using an iterative decomposition method as shown in the Appendix. To this end, we formulate a prior density of the scattering coefficient. According to the sparsity of the environment, we model  $\bf x$ to follow a Bernoulli-Gaussian distribution in a limited interval, defined as
\begin{align}
  {p_{\rm x}}\left( {x_{n_{\rm s}}|{\bf{p}}} \right) & = ( {1 - \alpha + \lambda} )\delta \left( x_{n_{\rm s}} \right) \nonumber \\ 
  & + \alpha  {\cal N}\left( {x_{n_{\rm s}}|\theta^{\rm x} ,{\sigma ^{\rm{x}}}} \right)\big(u({x_{n_{\rm s}}}) - u({x_{n_{\rm s}}}-1)\big),
\end{align}
where ${\bf{p}} \buildrel \Delta \over = \left[ {\lambda, \alpha, \theta^{\rm x}, {\sigma ^{\rm{x}}}} \right]$ denotes all parameters of the Bernoulli-Gaussian distribution, $\delta \left(  \cdot  \right)$ is the Dirac function, $u\left(  \cdot  \right)$ is the step function,  $\alpha  $ is the sparsity coefficient of the environment, and $\lambda {\rm{ = }}\int_{x \in \left( { - \infty ,0} \right] \cup \left[ {1, + \infty } \right)} {\alpha {\cal N}\left( {x|\theta^{\rm{x}} ,{\sigma^{\rm{x}}}} \right)} {\rm d}x$. In addition, $\theta^{\rm x}  \in \left[ {{\rm{0}},{\rm{1}}} \right]$ and ${\sigma ^{\rm{x}}}$ represent the mean and the standard deviation of the environmental information $\bf x$, respectively.


The computational complexity of the proposed method is expressed as $\mathcal{O}\left(N_{\rm{s}}M\right)$. 
It is evident that the number of pixels, $N_{\rm{s}}$, significantly influences the computational complexity, as the number of pixels is much larger than the number of sensing resources, such as antennas and carriers. The proposed method can use pixels with large areas, thereby reducing the number of pixels and the resources required for sensing. The integral in \eqref{eqtx} needs to be calculated by numerical integration. However, since the integral of each pixel can be parallelized and the calculation can be completed in advance of algorithm execution, the impact on the overall computational complexity is relatively minor.

\section{Performance Analysis}\label{sec4}
\begin{figure}
    \centering
    \includegraphics[width=0.99\linewidth]{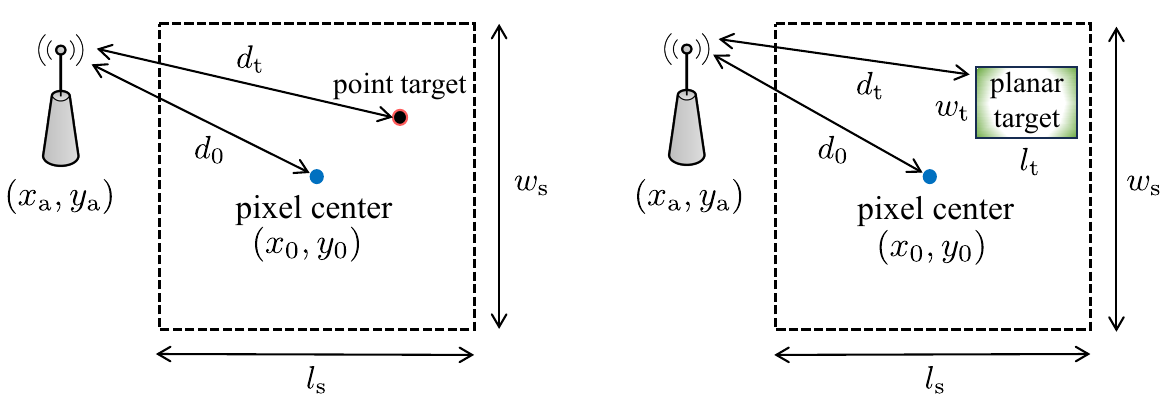}
    \caption{Two cases of pixel division errors caused by incomplete target filling. (left: point target case; right: planar target case).}
    \label{fig2}
\end{figure}

In this section, we analyze the impact of pixel division on sensing error and validate the effectiveness of the proposed method. As shown in Fig.~\ref{fig2}, we analyze the errors introduced by pixel division in two cases:  \textit{point targets} and  \textit{planar targets}. Let the antenna be located at $(x_{\rm a},y_{\rm a})$ and the central of the pixel at $(x_{\rm 0},y_{\rm 0})$ and introduce  $d(x,y) = \sqrt{(x_{\rm a} - x)^2 + (y_{\rm a} - y)^2}$ as the distance between the antenna and an arbitrary point in the pixel. We will use $d(x,y)$, $d(u,v)$  or $d(\tilde{u},\tilde{v})$ with identical meaning but different dummy variables. 

\subsection{Point Targets}

\subsubsection{Conventional Model}
For the point target shown in \ref{fig2}(a), modeled using conventional computational imaging as in \eqref{eq1}, the phase error $e_{1, \rm c}$ caused by the point target deviating from the pixel center, is derived by calculating the average error
\begin{align}\label{e1c}
  e_{1, \rm c} &= \int_{x_0 - l_{\rm s}/2}^{x_0 + l_{\rm s}/2}\int_{y_0 - w_{\rm s}/2}^{y_0 + w_{\rm s}/2}\frac{2\pi|d(x,y) - d_0 |}{l_{\rm s}w_{\rm s}\lambda}{\rm d}x{\rm d}y, 
\end{align}
where $d_0 = \sqrt{(x_{\rm a} - x_0)^2 + (y_{\rm a} - y_0)^2}$ is the distance between the antenna and the pixel. 
\subsubsection{Proposed Model}
The phase error $e_{1, \rm p}$ caused by the proposed method is expressed as
\begin{align}
  e_{1, \rm p} &= \int_{x_0 - l_{\rm s}/2}^{x_0 + l_{\rm s}/2}\int_{y_0 - w_{\rm s}/2}^{y_0 + w_{\rm s}/2}\frac{2\pi|d(x,y) - d_{\rm p} |}{l_{\rm s}w_{\rm s}\lambda}{\rm d}x{\rm d}y, 
\end{align}
where $d_{\rm p}$ is the average distance from pixel to antenna in an integrated form
\begin{align}\label{eqdp}
  d_{\rm p} &= \int_{x_0 - l_{\rm s}/2}^{x_0 + l_{\rm s}/2}\int_{y_0 - w_{\rm s}/2}^{y_0 + w_{\rm s}/2} \frac{d(u,v)}{l_{\rm s}w_{\rm s}}{\rm d}u{\rm d}v.  
\end{align}

Compared to the conventional computational imaging method, the proposed method has similar errors for point targets. This is because, for point targets, the average distance error caused by random distribution within pixels is similar to the average distance error obtained by the proposed integration method.
For square pixels, the average distance $d_{\rm p}$ calculated in \eqref{eqdp} is almost the same as the distance $d_{\rm 0}$ from the center of the pixel to the antenna in \eqref{e1c}. However, ideal point targets are unusual in practical scenarios. In a practical scenario, smaller targets can be accurately sensed by dividing smaller pixels until the resolution limit is reached, which is determined by the wavelength.

\subsection{Planar Targets}

\subsubsection{Conventional Model}
For the planar target as shown in Fig. \ref{fig2}(b) , the phase error $e_{2, \rm c}$ caused by the conventional method is expressed as
\begin{align}
  e_{2, \rm c} &= \int_{x_0 - l_{\rm s}/2}^{x_0 + l_{\rm s}/2}\int_{y_0 - w_{\rm s}/2}^{y_0 + w_{\rm s}/2}\frac{2\pi|d_{\rm t}(x,y) - d_0 |}{l_{\rm s}w_{\rm s}\lambda}{\rm d}x{\rm d}y, \label{eq20}\\
  d_{\rm t}(x,y) &= \int_{x - l_{\rm t}/2}^{x + l_{\rm t}/2}\int_{y - w_{\rm t}/2}^{y + w_{\rm t}/2} \frac{d(u,v)}{l_{\rm t}w_{\rm t}}{\rm d}u{\rm d}v,  
\end{align}
where $d_0 = \sqrt{(x_{\rm a} - x_0)^2 + (y_{\rm a} - y_0)^2}$ is the distance between the antenna and the pixel, and the length and the width of the planar target are denoted by $l_{\rm t}$ and $w_{\rm t}$. 

\subsubsection{Proposed Model}
The phase error $e_{2, \rm p}$ caused by the proposed method is expressed as
\begin{align}
  e_{2, \rm p} &= \int_{x_0 - l_{\rm s}/2}^{x_0 + l_{\rm s}/2}\int_{y_0 - w_{\rm s}/2}^{y_0 + w_{\rm s}/2}\frac{2\pi|d_{\rm t}(x,y) - d_{\rm p} |}{l_{\rm s}w_{\rm s}\lambda}{\rm d}x{\rm d}y,\label{eq23}\\
  d_{\rm p} &= \int_{x_0 - l_{\rm s}/2}^{x_0 + l_{\rm s}/2}\int_{y_0 - w_{\rm s}/2}^{y_0 + w_{\rm s}/2} \frac{d(\tilde{u},\tilde{v})}{l_{\rm s}w_{\rm s}}{\rm d}{\tilde u}{\rm d}{\tilde v}. 
\end{align}

For planar targets, there are two cases: when the planar target does not completely fill the pixel, and when the planar target completely fills the pixel. Incompletely filled pixels are usually at the edges of planar objects, while fully filled pixels are usually within planar objects. The error in both cases can be calculated using \eqref{eq23}. The proposed method effectively cancels the phase errors for planar targets. In \eqref{eq20}, there is a certain difference between $d_{\rm 0}$ calculated by the conventional algorithm and $d_{\rm t}$ in the planar target model. The $d_{\rm p}$ in \eqref{eq23} of the proposed model is close to $d_{\rm t}$ and makes the error as small as possible, especially when the planar target completely fills the pixel. The relevant integration results are shown in Fig. \ref{fig5}.

\begin{figure*}
  \centering
  \begin{minipage}[t]{0.98\textwidth}
      \centering
      \subfigure[Pixel size $l_{\rm s} = w_{\rm s} = 0.001 \rm m$, ROI size $L = W = 0.03 {\rm m}$.]{
      \includegraphics[height=3.2cm]{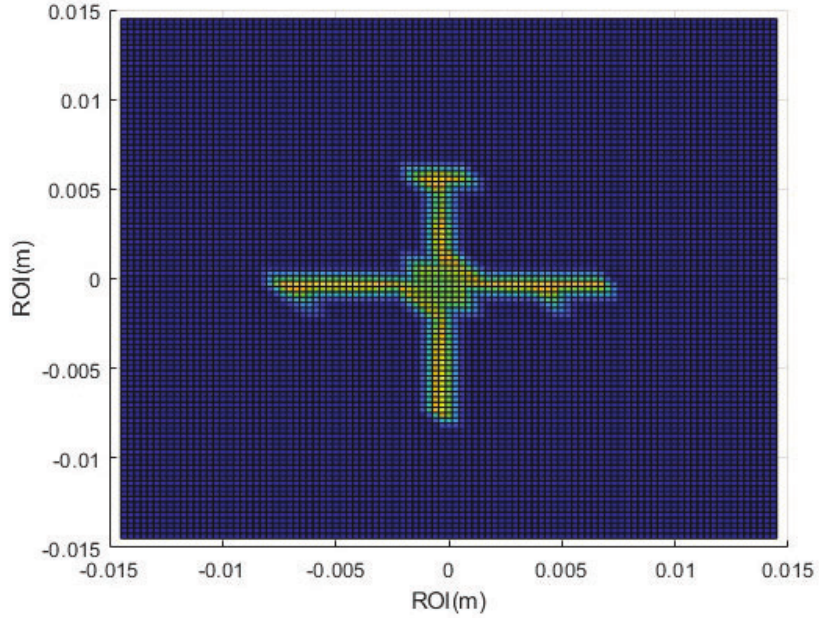}}
      \subfigure[Pixel size $l_{\rm s} = w_{\rm s} = 0.01 \rm m$, ROI size $L = W = 0.3 {\rm m}$.]{
      \includegraphics[height=3.2cm]{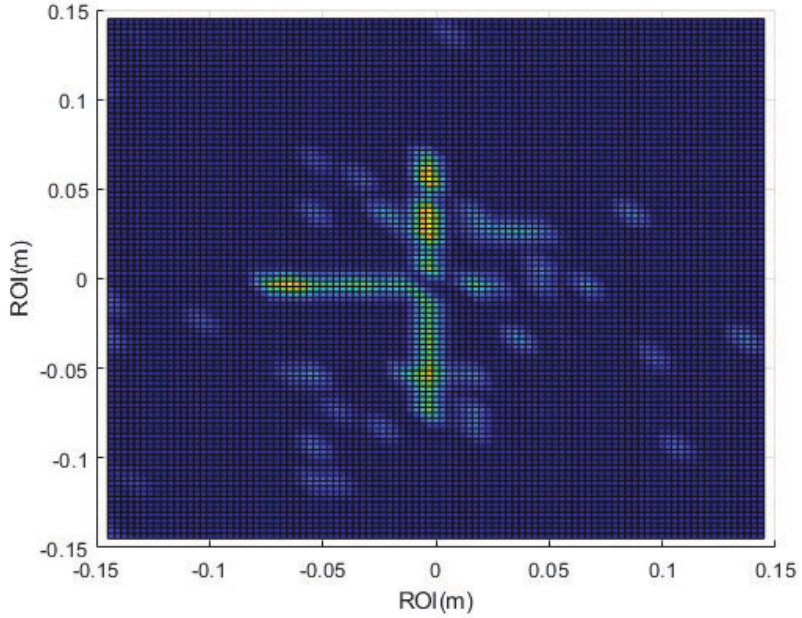}}
      \subfigure[Pixel size $l_{\rm s} = w_{\rm s} = 0.1 \rm m$, ROI size $L = W = 3 {\rm m}$.]{
      \includegraphics[height=3.2cm]{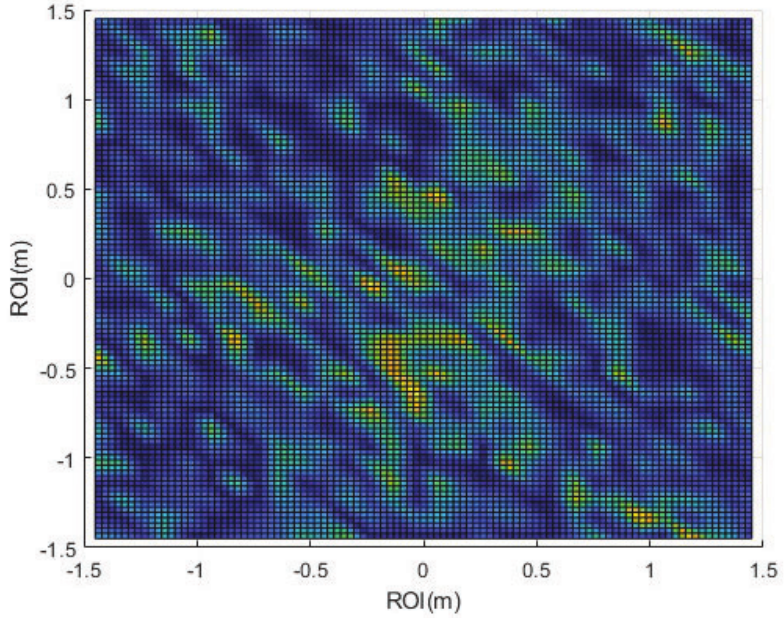}}
      \subfigure[Pixel size $l_{\rm s} = w_{\rm s} = 1 \rm m$, ROI size $L = W = 30 {\rm m}$.]{
      \includegraphics[height=3.2cm]{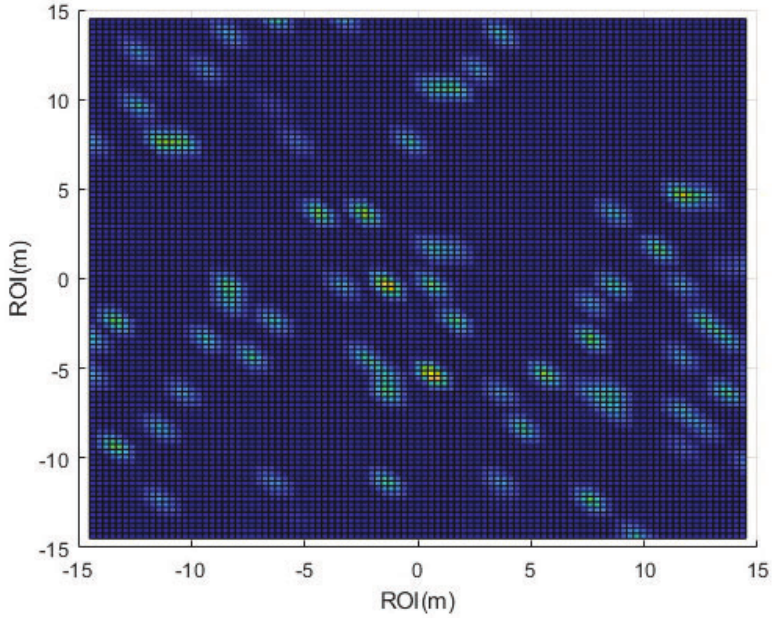}}

      \centering
      \subfigure[Pixel size $l_{\rm s} = w_{\rm s} = 0.001 \rm m$, ROI size $L = W = 0.03 {\rm m}$.]{
      \includegraphics[height=3.2cm]{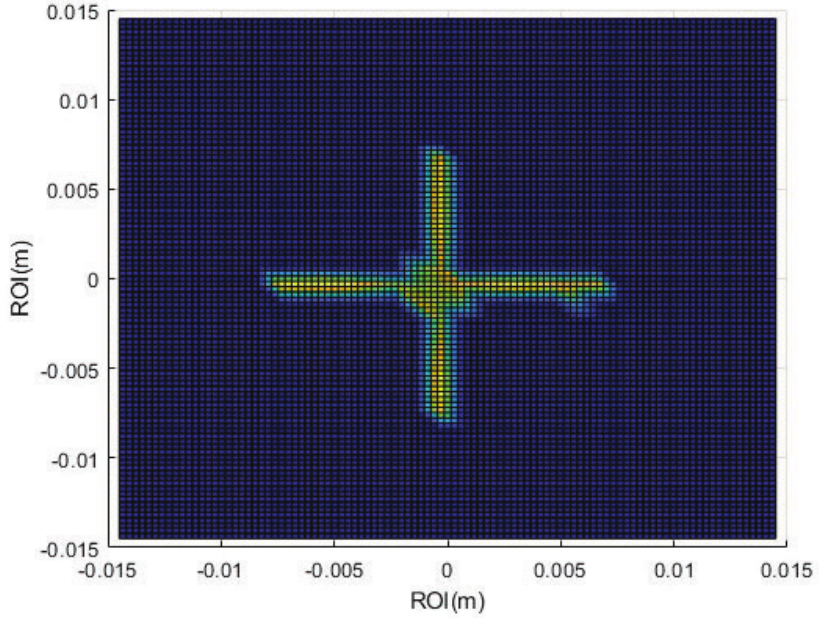}}
      \subfigure[Pixel size $l_{\rm s} = w_{\rm s} = 0.01 \rm m$, ROI size $L = W = 0.3 {\rm m}$.]{
      \includegraphics[height=3.2cm]{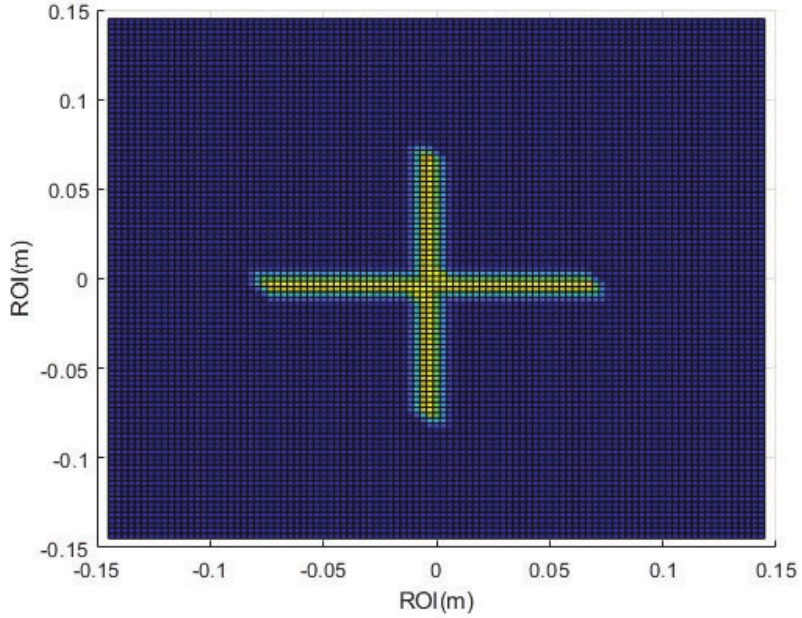}}
      \subfigure[Pixel size $l_{\rm s} = w_{\rm s} = 0.1 \rm m$, ROI size $L = W = 3 {\rm m}$.]{
      \includegraphics[height=3.2cm]{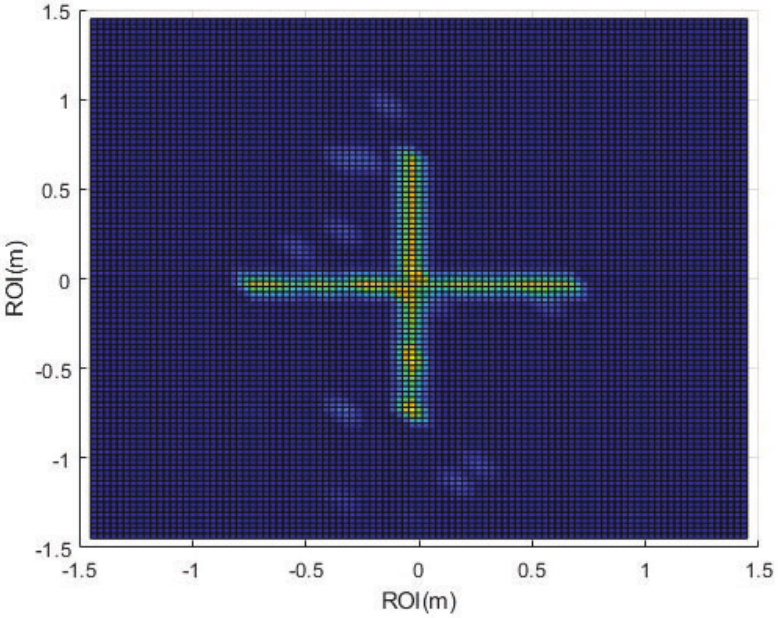}}
      \subfigure[Pixel size $l_{\rm s} = w_{\rm s} = 1 \rm m$, ROI size $L = W = 30 {\rm m}$.]{
      \includegraphics[height=3.2cm]{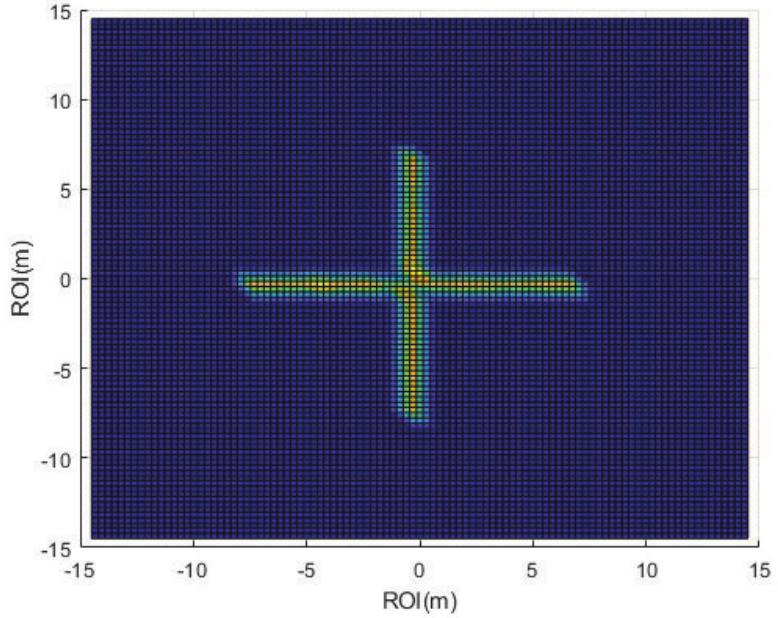}}
      \caption{The comparison sensing results between the baseline method (top row) and proposed method (bottom row).}
      \label{fig3}
  \end{minipage}
\end{figure*}

\section{Numerical Results}\label{sec5}
In this section, we describe the simulation environment, and present the numerical results for the proposed method, using the conventional computational imaging approach as a baseline for comparison.
\subsection{Simulation Environment and Metrics}
We consider a 2D scenario with a square ROI containing several scatters.
We use the ray tracing method proposed in \cite{xing} to generate received signals and propagation gains. The system parameters are set to $N_{\rm T} = N_{\rm R} = 20$ and $N_{\rm s} = 30\times 30$. The carrier frequency is set to $\rm 30$~GHz, and the signal-to-noise ratio (SNR) is 20~dB. Missed detection (MD) and false alarm (FA) rates are used to measure the sensing accuracy. We normalize the scatterer coefficient and set the threshold as 0.5 to decide whether there is a scatterer in the pixel or not. The original target object setting and the sensing results are compared to obtain MD and FA.

\subsection{Results and Discussion}
The intuitive sensing results of the baseline and the proposed method are compared in Fig.~\ref{fig3}, where a cross target is deployed in the ROI, and the pixel color represents the corresponding scattering rate for each pixel. The size of pixels is set to 0.001~m, 0.01~m, 0.1~m, and 1~m, respectively. It can be observed that the increase in pixel size significantly affects the accuracy of computational imaging, but the proposed method achieves computational imaging even under large-size pixels. 

\begin{figure*}
  \centering
  \begin{minipage}[t]{0.49\textwidth}
      \centering
      \subfigure[The relationship between the pixel size and MD.]{
      \includegraphics[width=0.92\textwidth]{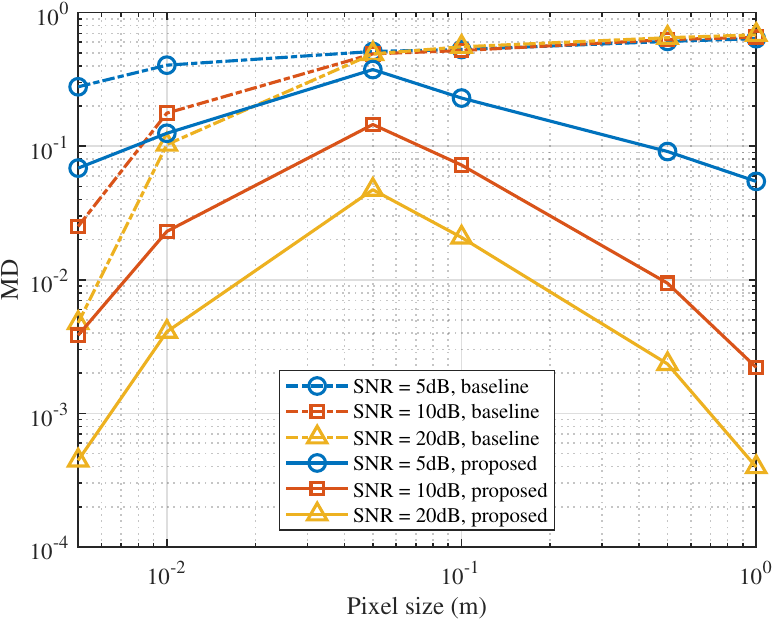}}
\end{minipage}
\begin{minipage}[t]{0.49\textwidth}
      \subfigure[The relationship between the pixel size and FA.]{
      \includegraphics[width=0.92\textwidth]{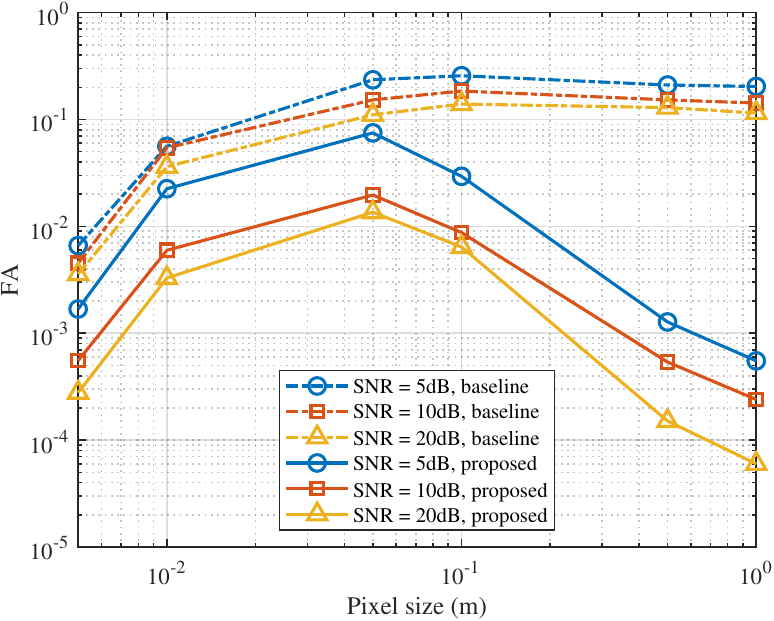}}
\end{minipage}
      \caption{The relationship between the pixel size and environment sensing performance.} \vspace{-0.5cm}
      \label{fig4}
\end{figure*}

\begin{figure}
  \centering
  \includegraphics[width=0.44\textwidth]{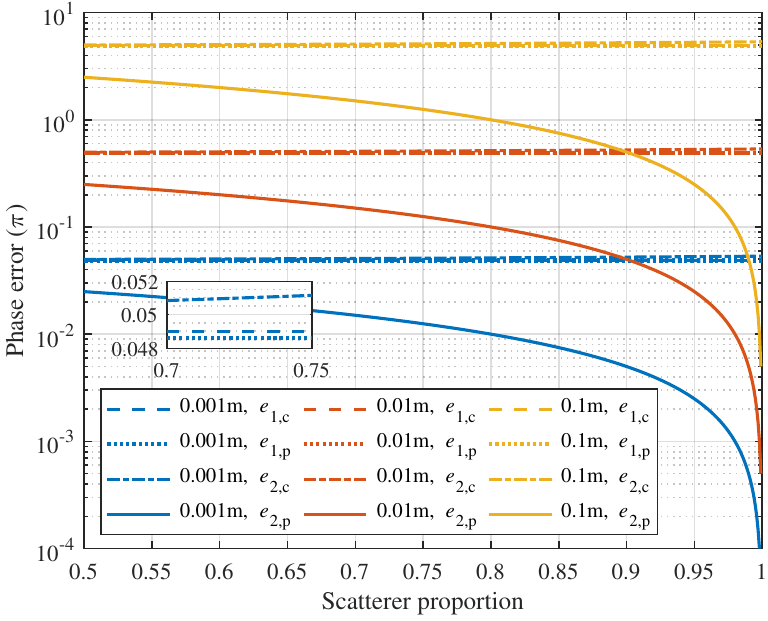}
  \caption{The relationship between the scatterer proportion and phase error.}
  \label{fig5}
  \end{figure}

Fig. \ref{fig4} shows the environment sensing results for different pixel sizes. When the pixel size is small (comparable to the wavelength), increasing the pixel size introduces greater phase errors, causing a gradual decline in imaging performance. When the pixel size further increases and the phase error is too large, conventional computational imaging cannot work, but the proposed method can still achieve accurate imaging. Additionally, based on the fundamental principles of computational imaging, the large-size pixel division reduces the correlation of the CS measurement matrix, leading to improved imaging performance and reduced MD and FA. Fig.~\ref{fig5} shows the numerical integration results for phase error of point targets and planar targets. As analyzed in Section~\ref{sec4}, the proposed method significantly cancels the phase errors. Especially in the case of planar targets, as the proportion of scatterers $(l_{\rm t}w_{\rm t})/(l_{\rm s}w_{\rm s})$ within a pixel increases, the phase error decreases, enabling accurate computational imaging.

\section{Conclusion}
In this paper, we proposed a novel computational imaging model based on the integral techniques to achieve the cancellation of the phase error, which can extend computational imaging to large-scale wireless communication scenarios. The proposed method leverages the continuous characteristics of object surfaces within the environment to facilitate large-scale environment sensing. Performance analysis and numerical results validate the effectiveness of the proposed method.

\appendix[GAMP algorithm]
The GAMP algorithm is summarized in Algorithm~1. The GAMP algorithm has defined two parameterized functions ${g_{\rm{in}}}\left(  \cdot  \right)$ and ${g_{\rm{out}}}\left(  \cdot  \right)$ which perform the generalization of the input and output variables. 
The parameterized functions ${g_{\rm{in}}}\left(  \cdot  \right)$, ${g_{\rm{out}}}\left(  \cdot  \right)$ and their derivatives ${g'_{\rm{in}}}\left(  \cdot  \right)$,  ${g'_{\rm{out}}}\left(  \cdot  \right)$ are shown as follows. To achieve the maximum posterior probability (MAP) estimation, the input function \cite{Rangan} is expressed as
\begin{equation}
{g_{\rm{in}}}\left( {\hat u,{\sigma ^{\rm{u}}},{\bf{p}}} \right) = \arg \mathop {\max }\limits_x {F_{\rm{in}}}\left( {x,\hat u,{\sigma ^{\rm{u}}},{\bf{p}}} \right),\label{gin}
\end{equation}
\begin{equation}
{F_{\rm{in}}}\left( {x,\hat u,{\sigma ^{\rm{u}}},{\bf{p}}} \right) = \log {p_{\rm x}}\left( {x|{\bf{p}}} \right) - \frac{1}{{2{\sigma ^{\rm{u}}}}}{\left( {\hat u - x} \right)^2},
\end{equation}
\begin{equation}
{g'_{\rm{in}}}\left( {\hat u,{\sigma ^{\rm{u}}},{\bf{p}}} \right) = \frac{1}{{1 - {\sigma ^{\rm{u}}}\frac{\partial ^2}{\partial  x^2} {\rm log}\left[{p_{\rm x}}\left( {x|{\bf{p}}} \right)\right]}},
\end{equation}
and the output function is expressed as
\begin{equation}
{g_{\rm{out}}}\left( {y,\hat q,{\sigma ^{\rm{z}}}} \right) = \frac{{y - \hat q}}{{{\sigma ^{\rm{w}}}} + {\sigma ^{\rm{z}}}},\label{gout}
\end{equation}
\begin{equation}
{g'_{\rm{out}}}\left( {y,\hat q,{\sigma ^{\rm{z}}}} \right) =  - \frac{1}{{{\sigma ^{\rm{w}}}} + {\sigma ^{\rm{z}}}}.
\end{equation}

{\small
\begin{algorithm}
  \caption{The GAMP Algorithm \cite{Rangan}}
  \label{alg1}
  \begin{algorithmic}[1]
  \REQUIRE
  Given the estimated propagation gain ${{\bf \hat H} \in {\mathbb{C}^{{ M } \times {1}}}}$ and the measurement matrix ${\bf{A }} \in {\mathbb{C}^{{M } \times {N_{\rm s} }}}$.
  \STATE
  \textbf{Initialization}: Set environmental information prior parameter $\bf{p}$. Defined ${g_{\rm{in}}}\left(  \cdot  \right)$ and ${g_{\rm{out}}}\left(  \cdot  \right)$ from (\ref{gin}), (\ref{gout}). Set $i = 0$, ${{\hat s}}\left( { - 1} \right) = 0$, ${{\hat z}_{{m }}}(i) = 0$,  ${\hat x_{{n_{\rm s} }}}\left( {{i}} \right) > 0$, $\sigma _{{n_{\rm s} }}^{\rm{x}}\left( {{i}} \right) > 0$.

  \WHILE {$\sum\limits_{{m }} {| {{{\bf \hat H}_{{m }}} - {{\hat z}_{{m }}}\left( {{i}} \right)} |}  > \varepsilon_{i} $, where $\varepsilon_{i} $ is a given error tolerance value in $i$-th iteration}
  \STATE
  For each $m$:

  {\small
  $\sigma _{{m }}^{\rm{z}}\left( {{i}} \right) = \sum\limits_{{n_{\rm s} }} {A _{{m },{n_{\rm s} }}^2} \sigma _{{n_{\rm s} }}^{\rm{x}}\left( {{i}} \right),$

  ${\hat q_{m }}\left( {i} \right) = \sum\limits_{n_{\rm s} } {A_{{m },{n_{\rm s} }}}{{\hat x}_{n_{\rm s} }}\left( {i} \right) - \sigma_{m }^{\rm{z}}\left( t \right) {\hat s_{{m }}}\left( {{i} - 1} \right),$

  ${\hat z_{{m }}}\left( {{i}} \right){\rm{ = }}\sum\limits_{{n_{\rm s} }} {{A_{{m },{n_{\rm s} }}}} {\hat x_{{n_{\rm s} }}}\left( {{i}} \right).$}
  \STATE
  For each $m$:

  {\small
  ${\hat s_{{m }}}\left( {{i}} \right) = {g_{\rm{out}}}\left( {{y_{{m }}},{{\hat q}_{{m }}}\left( {{i}} \right),\sigma _{{m }}^{\rm{z}}\left( {{i}} \right)} \right),$

  $\sigma _{{m }}^{\rm{s}}\left( {{i}} \right) =  - {g'_{\rm{out}}}\left( {{y_{{m }}},{{\hat q}_{{m }}}\left( {{i}} \right),\sigma _{{m }}^{\rm{z}}\left( {{i}} \right)} \right).$}
  \STATE
  For each $n_{\rm s} $:

  {\small
  $\sigma _{{n_{\rm s} }}^{\rm{u}}\left( {{i}} \right) = {\left[ {\sum\limits_{{n_{\rm s} }} {A _{{m },{n_{\rm s} }}^2\sigma _{{n_{\rm s} }}^{\rm{s}}\left( {{i}} \right)} } \right]^{ - 1}},$

  ${\hat u_{{n_{\rm s} }}}\left( {{i}} \right) = {\hat x_{{n_{\rm s} }}}\left( {{i}} \right) + \sigma _{{n_{\rm s} }}^{\rm{u}}\left( {{i}} \right)\sum\limits_{{m }} {{A _{{m },{n_{\rm s} }}}{{\hat s}_{{m }}}\left( {{i}} \right)}.$}
  \STATE
  For each $n_{\rm s} $:

  {\small
  ${\hat x_{{n_{\rm s} }}}\left( {{i}{{ + 1}}} \right) = {g_{\rm{in}}}\left( {{{\hat u}_{{n_{\rm s} }}}\left( {{i}} \right),\sigma _{{n_{\rm s} }}^{\rm{u}}\left( {{i}} \right),{\bf{p}}} \right),$

  $\sigma _{{n_{\rm s} }}^{\rm{x}}\left( {{i}{{ + 1}}} \right) = \sigma _{{n_{\rm s} }}^{\rm{u}}\left( {{i}} \right){g'_{\rm{in}}}\left( {{{\hat u}_{{n_{\rm s} }}}\left( {{i}} \right),\sigma _{{n_{\rm s} }}^r\left( {{i}} \right),{\bf{p}}} \right).$}
  \STATE

  {\small
  ${i} = {i} + 1.$}
  \ENDWHILE
  \ENSURE
  Estimated sparse environment target object ${\hat x_{{n_{\rm s} }}}\left( {{i}} \right)$.
  \end{algorithmic}
  \end{algorithm}
}

\ifCLASSOPTIONcaptionsoff
  \newpage
\fi

\bibliographystyle{IEEEbib}
\bibliography{ref}
\end{document}